\documentclass[aps,prd,preprint, floatfix,preprintnumbers,nofootinbib,superscriptaddress]{revtex4}
\usepackage[utf8]{inputenc}
\usepackage{fancybox}
\usepackage{hhline}
\usepackage{dcolumn}
\usepackage{textcomp}
\usepackage{epsfig,graphics,graphicx}
\usepackage{amsfonts,amssymb,amsmath}
\usepackage{pifont}
\usepackage{bm}
\usepackage{longtable} 
\usepackage{appendix}
\usepackage{lscape}
\usepackage[mathscr]{euscript}
\usepackage{mathrsfs}
\usepackage{multirow}
\usepackage{rotating}
\usepackage{color}

\newcommand{\beq}{\begin{equation}}
\newcommand{\eeq}{\end{equation}}
\newcommand{\bea}{\begin{eqnarray}}
\newcommand{\eea}{\end{eqnarray}}

\newcommand{\ord}[1]{{\cal{O}}( #1 )}
\newcommand{\B}{{\bf B}}
\newcommand{\Bdag}{{\bf B^\dagger}}

\usepackage{sansmath}
\usepackage{lmodern}
\newcommand{\gA}{\mathring g_A}
\newcommand{\BChPTNc}{$BChPT\times{ 1/N_c}~$}

\DeclareFontFamily{OT1}{pzc}{}
\DeclareFontShape{OT1}{pzc}{m}{it}%
              {<-> s * [0.900] pzcmi7t}{}
\DeclareMathAlphabet{\mathpzc}{OT1}{pzc}%
                                 {m}{it}
\DeclareMathAlphabet{\mathcalligra}{T1}{calligra}{m}{n}
\usepackage{hyperref}
\hypersetup{pdfauthor={me}, colorlinks=true, citecolor=blue, urlcolor=blue, linkcolor=black}
\begin{document}
\preprint{\vbox{\hbox{ JLAB-THY-19-2915} }}
\title{\phantom{x}
\vspace{-0.5cm}     }
\title{\centering   Baryon $\boldsymbol {\sigma}$ terms   in SU(3)  BChPT$\times$1/N${  \rm  \sansmath_c}$ }
\author{Ishara P. Fernando}
\thanks{Speaker.  9th International workshop on Chiral Dynamics,
		17-21 September 2018,
		Durham, NC, USA }
\affiliation{	Hampton University \& Thomas Nelson Community College,  Hampton , VA, USA}
\email{ishara@jlab.org}
\author{Jos\'{e} L. Goity}
\affiliation{	Hampton University,  Hampton , VA  \& Jefferson Lab, Newport News, VA, USA}
\email{goity@jlab.org}
\begin{abstract}
ChPT and the $1/N_c$ expansion provide systematic frameworks for  the strong interactions at low energy. A combined framework of both expansions has been developed and applied for baryons with three light-quark-flavors. The small scale expansion of the combined approach is identified as the $\xi$-expansion, in which the power counting of  the expansions is linked according to $\ord{p}=\ord{1/N_c}=\ord{\xi}$. The physical  baryon masses as well as   lattice QCD baryon masses for  different quark mass masses are analyzed to $\ord{\xi^3}$ in that framework.
$\sigma$ terms are  addressed using the Feynman Hellmann theorem.  For the nucleon,  a useful connection between the deviation of the Gell-Mann-Okubo relation and the $\sigma$ term  $\sigma_{8N}$ associated with the scalar density $\bar u u+\bar d d-2\bar s s$ is identified. In particular, the deviation from the tree level relation $\sigma_{8N}=\frac 13(2 m_N-m_\Sigma-m_\Xi)$, which gives rise to the so called $\sigma$-term puzzle,  is studied in the $\xi$-expansion. A large  correction  non-analytic in $\xi$  results for that relation, making plausible the resolution of the puzzle. 
Issues with the determination of the strangeness $\sigma$ terms are discussed, emphasizing the need for lattice calculations at smaller $m_s$ for better understanding the range of validity of the effective theory. 
The analysis presented here leads to $\sigma_{\pi N}=69(10)$~MeV  and $\sigma_{\pi \Delta}=60(10)$~MeV.
\end{abstract}
\maketitle
\section{Introduction }

Combining $BChPT$ and the $1/N_c$  expansion \cite{Jenkins:1995gc,FloresMendieta:1998ii,FloresMendieta:2012dn,CalleCordon:2012xz} in baryons with three light quark flavors leads to  an improvement  in the description  of  baryon masses and currents \cite{CalleCordon:2012xz,Fernando:2017yqd,Flores-Mendieta:2014vaa,FernandoGoityCD18-1}   to one-loop.  A link between the chiral and the 1/Nc expansions is necessary in order to establish an unambiguous power counting:  the counting where  $\ord{p}=\ord{1/N_c}=\ord{\xi}$, closely related to the  small scale expansion \cite{Hemmert:1996xg,Hemmert:1997ye}, is in practice the most effective one. 
In this framework, the effective Lagrangians to $\ord{\xi^3}$ can be found in Ref. \cite{Fernando:2017yqd,Fernando:2018jrz}.
The chiral Lagrangian relevant to the discussion of masses  up to $\ord{\xi^3}$ and including electromagnetic contributions is given by \cite{Fernando:2017yqd,Fernando:2018jrz}: 
\bea
{ \cal{L}}_\B^{\text{mass}} &=& \Bdag \left( iD_0 + \mathring{g}_A u^{ia}G^{ia}- \frac{C_{\text{HF}}}{N_c}\hat{S}^2 + \frac{c_1}{2\Lambda} \hat{\chi}_+  +
\frac{c_2}{\Lambda}\,  \chi^0_+ + \frac{c_3}{N_c\,\Lambda^3} \;\hat\chi_+^2 \right.  \\
& & \left. +\frac{h_1 \Lambda}{N_c^3} 
\hat{S}^4+ \frac{h_2}{N_c^2\Lambda} \hat\chi_+\hat{S}^2+ \frac{h_3}{N_c\Lambda} \chi^0_+ \hat{S}^2+\frac{h_4}{N_c\, \Lambda} \; \chi_+^a\{S^i,G^{ia}\} +\alpha \hat{Q} + \beta \hat{Q}^2\right) \B ,\nonumber
\label{eq:Mass_Lagrangian}
\eea
where $\mathring{g}_A$ is the  axial coupling constant   identified at LO with  $\frac{6}{5}g_A^N$,  where   $g_A^N=1.2724(23)$. The low energy constants (LECs) $C_{\text{HF}}$, $c_{1-4}$, $h_{1-4}$ and $\alpha, \beta$ can be fixed \cite{Fernando:2017yqd} by fitting the  baryon masses to the experimental data and  to results from lattice QCD (LQCD) calculations \cite{Alexandrou:2014sha} at varying quark masses. Using standard notation,  $\hat{\chi}_+=\tilde{\chi}_+ \thinspace + \thinspace \chi_+^0$,  where $\thinspace \chi_+^0 = \frac{1}{3} \text{Tr} \chi_+$, provide  the quark mass dependent terms.
$\hat{Q}$ is the electric charge operator. The electromagnetic contribution to the $p-n$ mass difference is $\alpha + \beta$, whereas the electromagnetic contribution to the Gell-Mann-Okubo (GMO) formula is $-\frac{4}{3}\beta$. Up to  $\ord{\xi^3}$ the baryon mass formula, neglecting isospin breaking,  reads:
\bea
m_\B &=& M_0 + \frac{C_{\text{HF}}}{N_c}\hat{S}^2 - \frac{c_1}{\Lambda} 2 B_0 \left( \sqrt{3} m_8 \hat Y  + N_c m_0\right) - \frac{c_2}{\Lambda} 4B_0 m_0 
 \nonumber \\
& & - \frac{h_1 \Lambda}{N_c^3} \hat S^4- \frac{h_2}{N_c^2\Lambda}  4 B_0  (\sqrt{3} m_8 \hat Y  + N_c m_0)\hat{S}^2 - \frac{h_3}{N_c\Lambda} 4B_0 m_0 \hat{S}^2   \\
& & - \frac{h_4}{N_c\, \Lambda}  \frac{4B_0m_8}{\sqrt{3}} \left( 3 \hat{I}^2 - \hat{S}^2 - \frac{1}{12}N_c (N_c+6) + \frac{1}{2}(N_c+2) \hat Y - \frac{3}{4} \hat Y^2 \right)
+ \delta m_B^{\text{loop}} , \nonumber
\label{eq:Baryon_Mass_formula}
\eea
where $M_0$ is the $\ord{N_c}$ spin-flavor singlet piece of the baryon masses, $\hat S$, $\hat I$ and $\hat Y$ are respectively the baryon spin, isospin and hypercharge operators,    the term proportional to $C_{HF}$   gives the LO hyperfine mass splittings between different spin baryons, and $m_0$ and $m_8$ are the singlet and octet components of the quark masses. $\delta m_B^{\text{loop}} $ gives the one-loop contributions 
   $\ord{\xi^2}$ and  $\ord{\xi^3}$. It is straightforward to generalize \ref{eq:Baryon_Mass_formula} to include isospin breaking.
In the following the definitions are used:  $m_0=\frac{1}{3}\left(2 \hat{m}+m_s\right)$,   $m_3=m_u-m_d$ and  $m_8=\frac{1}{\sqrt{3}}\left(\hat{m}-m_s\right)$,  where  $\hat{m}=\frac{1}{2}\left(m_u+m_d\right)$. More details on the self energy one-loop corrections obtained in \BChPTNc can be found in these proceedings \cite{FernandoGoityCD18-1}.

\section{$\sigma$-terms  }

The  matrix elements of scalar quark densities are of high interest. At zero momentum they are related via the Feynman-Hellmann theorem to the slope of the hadron mass with respect to the corresponding quark mass, \footnote{ Although obvious,   $\sigma$ terms,  being observable quantities, are  independent of the renormalization scheme used in QCD. The expression  \ref{eq:Feynman-Hellmann} normally used is valid in a mass independent   scheme such as $\overline {MS}$.}

\begin{eqnarray} \label{eq:Feynman-Hellmann}
\sigma_{f  \B}(m_f) &=& m_f \frac{\partial}{\partial m_f}m_{\B}=\frac{m_f}{2m_\B}\langle\B\mid \bar q_f q_f\mid \B\rangle ,
\end{eqnarray}
where $m_f$ is the  mass of the $f$ quark flavor  ($f=u,d,s$), the state $\mid \B\rangle$ is the physical state   for that quark mass and normalized according to  $\langle \B'\mid \B\rangle=(2\pi)^3 2 m_B \delta^3(\vec p'-\vec p)$,  and  $\sigma_{f  \B}$ is the corresponding $\sigma$ term. $\sigma$ terms for combinations of quark masses such as   $m_0$, $m_3$ and $m_8$  are defined in the same way. 
Empirical access to $\sigma$ terms is difficult in the case of baryons, being only possible for $\sigma_{\pi N}=\sigma_{(u+d)  N}(\hat m) $ via analysis of $\pi N$ scattering. In the case of other $\sigma$ terms it is clear that the necessary information will have to come from LQCD calculations, where tracing the baryon mass dependency with respect to quark masses is becoming increasingly accurate.
The actual contribution of a given quark flavor mass to the mass of the hadron, keeping the rest of the quark flavor masses fixed,  is then given by:
 \begin{equation}\label{eq:DeltamB}
 \Delta m^f_\B(m_{f})=\int_0^{m_f}  \frac{1}{\mu}\sigma_{f \B}(\mu) d\mu ,
 \end{equation}
  which in the limit of small $m_f$ coincides with the $\sigma$ term.

In this note, the focus is  on the determination of $\sigma_{\pi N}$ using the Feynman-Hellmann theorem and results for baryon masses in $SU(3)$, as presented in Ref. \cite{Fernando:2018jrz}, with additional brief discussions  of $\sigma$ terms of  $\Delta$ and hyperons, and the issue of    the quark mass dependence of $\sigma$ terms, namely  the range in $m_q$ where the effective theory may be trusted in their description.

\subsection{$ { {\sigma}_{ {\pi N}}}$ }{\label{addressing-puzzle}

The determination of $\sigma_{\pi N}$ has a long history spanning many decades. Its extraction from the analyses of $\pi N$ scattering has given values that range from 45 MeV \cite{Gasser:1988jt,Gasser:1990ce,Gasser:1990ap} to 64 MeV \cite{Pavan:2001wz,Alarcon:2011zs,Hoferichter:2015dsa,Hoferichter:2016ocj}, with the larger values being from more recent analyses where their increment with respect to the olg ones is understood to be a consequence of a change in the input $\pi N$ scattering lengths. From a practical use point of view, $\sigma_{\pi N}$  has become very important in the studies  of dark matter searches  \cite{Ellis:2018dmb} in the scenarios where dark matter has scalar couplings to quarks.

$\sigma_{\pi N}$ can be expressed by  the combination of $\sigma$ terms:
\begin{eqnarray} \label{eq:sigma-relation-1}
\sigma_{\pi N} &=& \hat{\sigma} + 2 \frac{\hat{m}}{m_s} \sigma_{s N},
\end{eqnarray} 
where $\hat \sigma=\sqrt{3}\frac{\hat m}{m_8} \sigma_{8 N}$.  To LO in quark masses $\sigma_{8 N}$ is given by a combination of octet baryon masses, namely:
\begin{eqnarray} \label{eq:sigma8-literature}
\sigma_{8 N}&=&\frac 13(2 m_N-m_\Sigma-m_\Xi),\\
 &=&\frac 19\left( \frac{5N_c-3}{2} m_N-(2N_c-3)  m_\Sigma-\frac{N_c+3}{2} m_\Xi \right) ~~~\text{for general $N_c$},\nonumber
\end{eqnarray}
which leads to $\hat \sigma\sim 25$ MeV. Since the contribution of the term proportional to $ \sigma_{s N}$, being OZI suppressed, should be expected to be small, at this lowest order in the quark masses there is a puzzle between the empirically obtained  values of $\sigma_{\pi N}$ and the relation $\sigma_{\pi N}\sim \hat \sigma$. Either the latter is badly broken, and/or the relation \ref{eq:sigma8-literature} has large corrections. It will be shown that the latter is the case.  It is  argued that the puzzle is further emphasized  by the observation that the Gell-Mann-Okubo relation \footnote{The GMO relation is defined by the mass combination: $3m_\Lambda+m_\Sigma-2(m_n+m_\Xi)$,   valid for all $N_c$.}
receives small deviations, and so it would be difficult to understand why  \ref{eq:sigma8-literature} should receive large corrections \cite{Leutwyler:2015jga}. Following Ref. \cite{Fernando:2017yqd}, and based on the $1/N_c$ expansion one finds  that the corrections to the GMO relation are suppressed by a factor $1/N_c$ at large $N_c$, while the corrections to the mass relation generalized in $N_c$ as shown in \ref{eq:sigma8-literature}   are $\ord{N_c}$.
The deviation from the GMO relation, $\Delta_{GMO}$, in the calculation to one-loop is independent of the NLO LECs and given solely by non-analytic finite contributions, which depend on $\gA/F_\pi$, $C_{HF}$ and the GB masses. The same is the case for the deviations from   \ref{eq:sigma8-literature}, denoted here by $\Delta \sigma_{8 N}$. Performing the analysis at generic $N_c$, one finds that $\Delta_{GMO}$ is indeed $\ord{1/N_c}$ at large $N_c$, and in terms of the $\xi$ power counting  it is $\ord{\xi^4}$ (an extra factor $1/N_c$ over the nominal $\ord{\xi^3}$ of the loop corrections), while $\Delta \sigma_{8 N}$  is $\ord{\xi^2}$ with a pre-factor $N_c$ \footnote{Note that $\sigma_{8 N}=\ord{N_c}$, while $\sigma_{sN}=\ord{N_c^0}$}. Thus they have entirely different behaviors, and on these grounds it is  entirely plausible  that  $\Delta \sigma_{8 N}$ can be as large as the resolution to the puzzle requires.  It is also observed that in the physical case the ratio $\Delta \sigma_{8 N}/\Delta_{GMO}\sim -14$, which is independent of $\gA$ and $F_\pi$, has only a small dependency on the LEC $C_{HF}$, and thus it is determined almost entirely in terms of the GB masses. Since the large corrections  $\Delta \sigma_{8 N}$ are due to the rather large value of $m_s$, it is important  to check how  $\sigma_{8 N}$ is as a function of $M_K$. This is shown in Figure 1, which clearly illustrates the following point: the non-analytic contributions to $\sigma_{8 N}$ are not large (compare $\sigma_{8 N}$ with the tree contribution $\sigma_N^{tree}(\mu=m_\rho)$). The corrections to the mass combination denoted here by $\sigma_8^{rel}$ are very small, but they result  from  two large contributions, the $\sigma_N^{tree}(\mu=m_\rho)$ and a non-analytic one that largely   cancels it. Thus, a large entirely non-analytic correction   $\Delta \sigma_{8 N}$ is the result. The figure also shows the behavior of $\sigma_s$, which has a large relative variation in the displayed interval; its size is nonetheless  natural, leading in Eq. \ref {eq:sigma-relation-1} to a small contribution by that term  of the order of a few MeV. As discussed later, the $\sigma_s$ terms are outside of the range of validity of the effective theory for the physical $m_s$ values.
In order to check that the effective theory is giving reasonable results, one can make use of the calculated $\Delta_{GMO}$ and check with its actual value: as shown below, this works very well; even more, the octet baryons in the loop contribute 43\% of $\Delta_{GMO}$, thus  the contribution by the decuplet is crucial. One can also infer from $\Delta_{GMO}$ a value for the LO axial coupling $\gA$: it is about 20\% smaller than the physical one, in line with that obtained in the analysis of  axial couplings \cite{Fernando:2017yqd,FernandoGoityCD18-1}. If one only considers the contributions by the octet baryons, which is  itself $\ord{1/N_c}$, in order to obtain the physical $\Delta_{GMO}$ the $\gA$ needed must be larger,   conflicting  with the analysis of the axial  couplings \cite{Fernando:2017yqd,FernandoGoityCD18-1}.
 
 At this point, the effective theory can determine $\sigma_{8 N}$ from  \ref{eq:sigma8-literature} and the calculated $\Delta \sigma_{8 N}$. To determine $\sigma_{\pi N}$ one needs further information on the baryon masses. That information is provided by LQCD, as for instance in the analysis of octet and decuplet masses of Ref. \cite{Alexandrou:2014sha}, where $m_s$ is kept approximately fixed and $\hat m$ is varied. A fit to the masses allows for a direct extraction of $\sigma_{\pi N}$ and also an estimate, albeit with large error, of $\sigma_{sN}$. As discussed below, the end result is that the relation $\sigma_{\pi N}\simeq \hat \sigma$ is approximately well satisfied. The most direct determination of $\sigma_{8 N}$ is thus carried out making use of the ratio $\Delta \sigma_{8 N}/\Delta_{GMO}$ using a value of $C_{HF}$ as obtained in the fit to octet and decuplet masses  and correcting $\Delta_{GMO}$ by         EM   and $m_u-m_d$ isospin breaking effects (see Ref. \cite{Fernando:2017yqd} for details), giving  $\hat \sigma\simeq 70$ MeV,  which leads to a value for $\sigma_{\pi N}$ which is at the upper range of values obtained in previous studies.

The question is up to what extent is the determination of $ \sigma_{8 N}$ discussed here realistic.  It is clear, as emphasized below, that the $\sigma$ terms associated with the strange quark at its physical mass cannot be described well by the effective theory. This implies also that the description of the hyperon masses in the physical case are somewhat outside the range where one can trust the effective theory. Thus, both the parameter free calculations of $\Delta_{GMO}$ and $\Delta \sigma_{8 N}$ may not be as accurate as one would wish. There is little doubt that the analysis presented here would work reliably for a smaller $m_s$, for $M_K<300-400$ MeV or so (see Fig. \ref{fig:1}). The only way this can be established is via LQCD calculations  with lighter $m_s$ than the ones presently available. Such calculations would indeed provide important additional insights on the $\sigma$ terms and more in general on the effectiveness  of the different versions of BChPT, in particular the present one, which would be greatly welcomed.

\begin{figure}[]
\begin{center}
\includegraphics[height=5.5cm,width=9cm,angle=0]{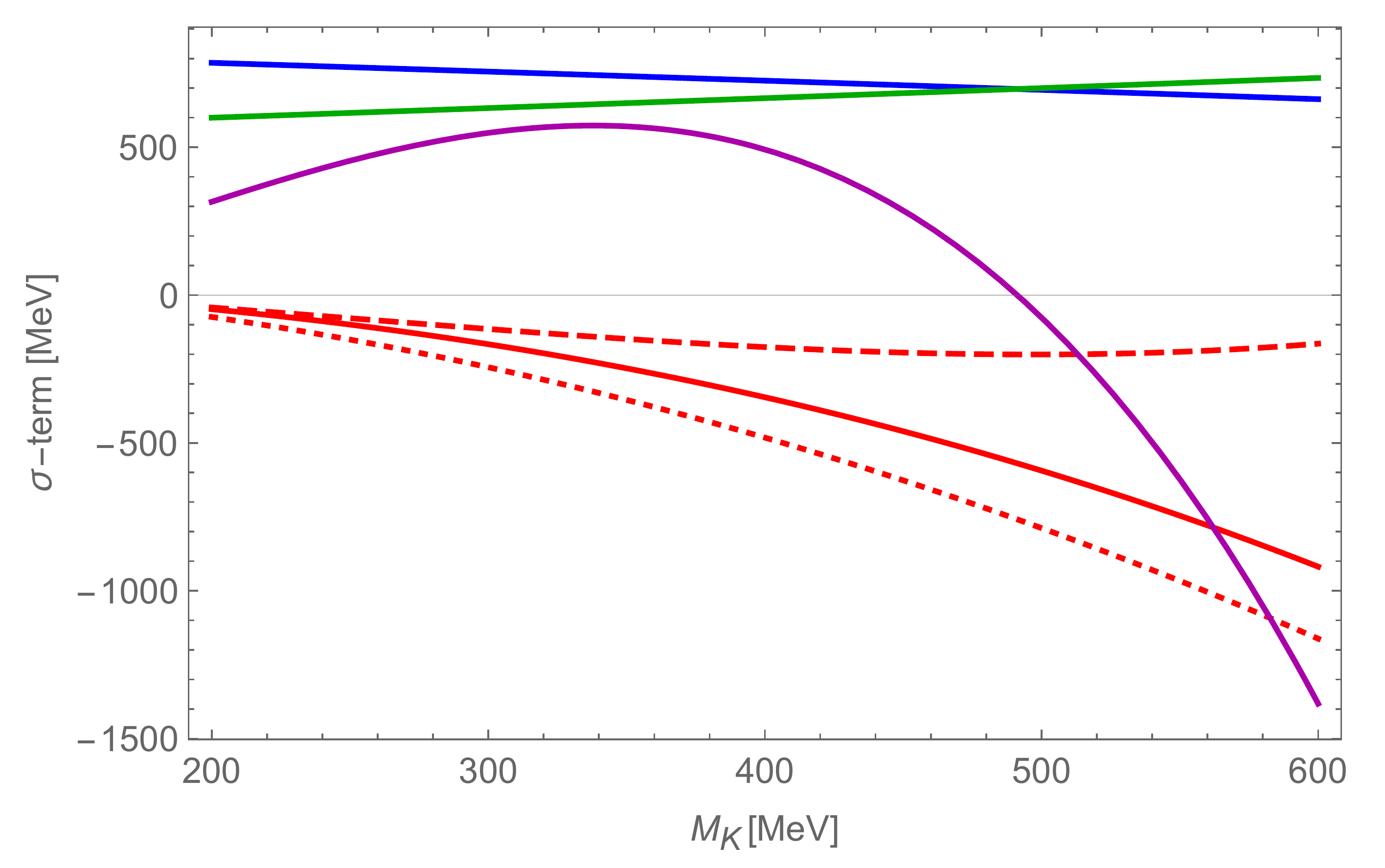}
\end{center}
\caption{$\sigma$ terms as a function of $M_K$ from baryon masses to $\ord{\xi^3}$. $\sigma_{8 N}$ full red, $\sigma_{8N}^{{\rm tree}}(\mu=m_\rho)$ short-dash red, $\sigma_{8 N}$ from the mass relation \ref{eq:sigma8-literature} dashed red, $10\times \sigma_{sN}$ purple, $10\times \hat \sigma_N$ green, and $10\times \sigma_{\pi N}$ blue. Based on the analysis of Ref. \cite{Fernando:2017yqd}.}
\label{fig:1}
\end{figure}

\subsection{Other $ \sigma$ terms }

A similar analysis to the case of the Nucleon can be carried out for the $\Delta$. In that case there is the following LO relation for $\sigma_\Delta ^8$:
\begin{eqnarray}\label{Eq:sigma10rel}
 \sigma_{8 \Delta}&=&\frac{N_c}{3} \left(m_\Delta-m_{\Sigma^*}\right) - \frac{5(N_c-3)}{12}\left(m_\Lambda-m_\Sigma \right)  ,
\end{eqnarray}
whose deviations at NLO are again calculable as in the case of the Nucleon.  Since in the large $N_c$ limit the $\Delta$ and Nucleon become degenerate, their respective $\sigma$ terms must also become identical up to terms sub-leading in $1/N_c$. That regime is however reached at very large $N_c$ (fixed $m_q$) for the contributions  non-analytic in $\sqrt{m_q} \times N_c$.

$\sigma$ terms satisfy the same tree level relations as baryon masses do. Indeed, the GMO, Equal spacing and the G\"ursey-Radicati (if the LEC $h_3$ is neglected)  mass relations, satisfied by  tree contributions   up to $\ord{\xi^3}$,  are automatically satisfied by the corresponding $\sigma$ terms. Since the non-analytic corrections to those relations are all $1/N_c$ suppressed,  the corresponding $\sigma$ term relations   have small deviations.   There are further tree level relations satisfied by $\sigma$ terms corresponding to different quark masses, in particular relating the $\sigma$ terms corresponding to $m_s$ with the $\hat m$ ones.  The  corrections to those are not $1/N_c$ suppressed and thus they  receive large non-analytic corrections. As shown later, the $\sigma_\B^s$ terms show significant curvature starting at $M_K\sim   250$ MeV, indicating the range where the effective theory can be trusted with their calculation. Those additional $\sigma$ terms may be of general  interest in LQCD calculations and  the corresponding  tests of the effective theory they can provide. 

\section{Results using LQCD inputs }
 In the analysis of Ref.  \cite{Fernando:2018jrz}, both physical  and  LQCD  baryon masses are considered.  The  LQCD baryon masses have been obtained for approximately fixed $M_K$,   varying $m_u=m_d$ in a range from the physical limit   up to  $M_\pi\sim 300$~MeV \cite{Alexandrou:2014sha}. Three different   fits were performed,  shown in the Table (\ref{Table:fits}),  which contains some additional results to those given  in \cite{Fernando:2018jrz}. 
 The ratio $\mathring{g}_A/F_{\pi}$ is also a fitting parameter for the first two fits  and it is consistent with the value extracted from $\Delta_{\text{GMO}}$ and also the one obtained from the analysis of axial couplings \cite{Fernando:2017yqd,FernandoGoityCD18-1}. The value of   $C_{\text{HF}}$ is determined  most accurately by the physical $\Delta-N$ mass splitting; its value obtained solely from the LQCD results is significantly different  and indication that the LQCD results do not determine accurately the hyperfine mass splittings,   extrapolating  to too small of a value at  the physical limit.
 For the physical case isospin breaking was taking into account, which allows to fix the EM coefficients $\alpha$ and $\beta$. For the present analysis, the importance of that correction is its effect on $\Delta_{GMO}$, whose value without EM is  that obtained with the physical masses plus $\frac 43 \beta$, a non-negligible effect of almost 3 MeV increase.

\begin{table}[h]
	\begin{center}\small
		\begin{tabular}{c@{\hspace{2pt}}c@{\hspace{2pt}}c@{\hspace{2pt}}c@{\hspace{2pt}}c@{\hspace{2pt}}c@{\hspace{2pt}}c@{\hspace{2pt}}c@{\hspace{2pt}}c@{\hspace{2pt}}c@{\hspace{2pt}}c}\\\hline\hline 
			&{\large$\frac{\mathring g_A}{F_\pi}$}&{\large$\frac{M_0}{N_c}$}&$C_{HF}$ & $c_1$ & $c_2$& $h_2$&$h_3$&$h_4$&$\alpha$&$\beta$\\[.21cm]
			Fit&${\rm MeV}^{-1}$ & MeV  & MeV & &&&&& MeV & MeV \\[5pt] \hline 
			1&$0.0126(2) $ &$364(1)$ & $166(23)$ &$-1.48(4) $&$ 0 $ &$ 0$&$0.67(9) $&$0.56(2)$&$-1.63(24) $&$2.16(22) $\\
			2 &$0.0126(3)$&$213(1)$&$179(20)$&$-1.49(4)$&$-1.02(5)$&$-0.018(20)$&$0.69(7)$&$0.56(2)$&$-1.62(24)$&$2.14(22)$\\
			3&$0.0126^*$&$262(30)$&$147(52)$&$-1.55(3)$&$-0.67(8)$&$0$&$0.64(3)$&$0.63(3)$&$-1.63^*$&$2.14^*$\\
			\hline\hline
			&   $\Delta^{\rm phys}_{GMO}$ &$\sigma_{8N}$ & $\Delta\sigma_{8N}$ &$\hat\sigma_N$ & $\sigma_{\pi N}$&$\sigma_{sN}$& $\sigma_{8\Delta} $ & $\Delta\sigma_{8\Delta} $ & $\hat{\sigma}_\Delta$ &   \\[.2cm]
			&${\rm MeV} $ & MeV  & MeV&MeV&MeV&MeV&MeV&MeV& MeV &  \\[5pt]\hline  
			1&$25.6(1.1) $ & $-583(24)$&$-382(13)$&$70(3)(6)$&$-$&$-$& $-496(46)$ & $-348(16)$ &$59(5)(6)$ &\\
			2 &$25.5(1.5)$&$-582(55)$&$-381(20)$&$70(7)(6)$&$69(8)(6)$&$-3(32)$& $-511(52)$ & $-352(22)$ & $60(10)(6)$& \\
			3&$25.8^*$ &$-615(80)$&$-384(2)$&$74(1)(6)$&$65(15)(6)$&$-121(15)$& $-469(26)$ & $350(27)$ & $56(4)(6)$ &\\
			\hline\hline
		\end{tabular}
	\end{center}
	\caption{ Results of  fits to baryon masses \cite{Fernando:2018jrz}.  Fit 1 uses only the physical octet and decuplet masses, Fit 2 uses the physical and the LQCD masses  from Ref. \cite{Alexandrou:2014sha}   with  $M_\pi\lesssim 300$ MeV, and Fit 3 uses only those LQCD masses and imposes the value of $\Delta^{\rm phys}_{GMO}$ determined by the physical masses (corrected in the calculation by the isospin breaking effects).  The renormalization scale $\mu$ and the scale $\Lambda$ are taken to be equal to $m_\rho$.   $^*$ indicates an input. A theoretical error of 6 MeV is estimated for $\hat \sigma$ and $\sigma_{\pi N}$.}
	\label{Table:fits}
\end{table}
\begin{figure}[h!!]
	\begin{center}
		\epsfig{file=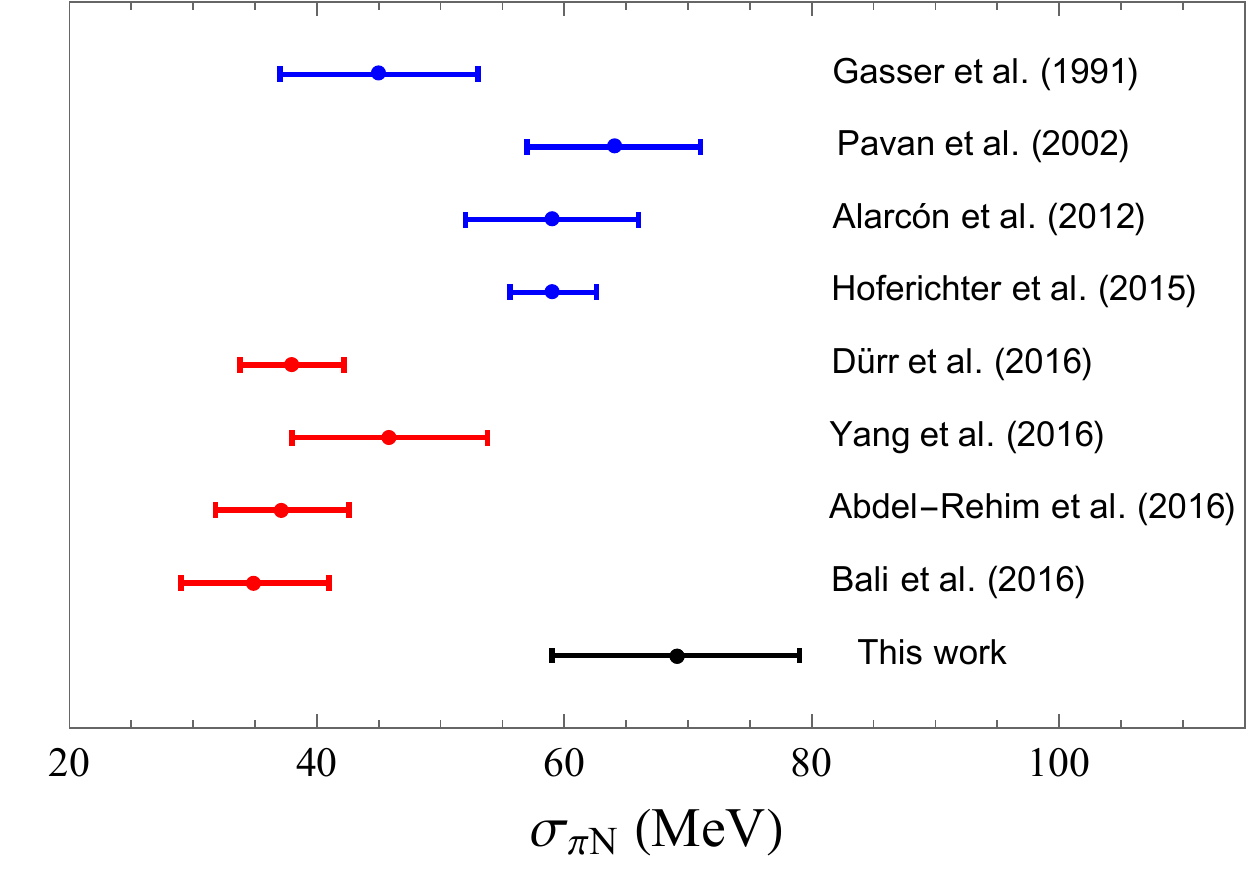,width=0.45\textwidth,angle=0} ~\epsfig{file=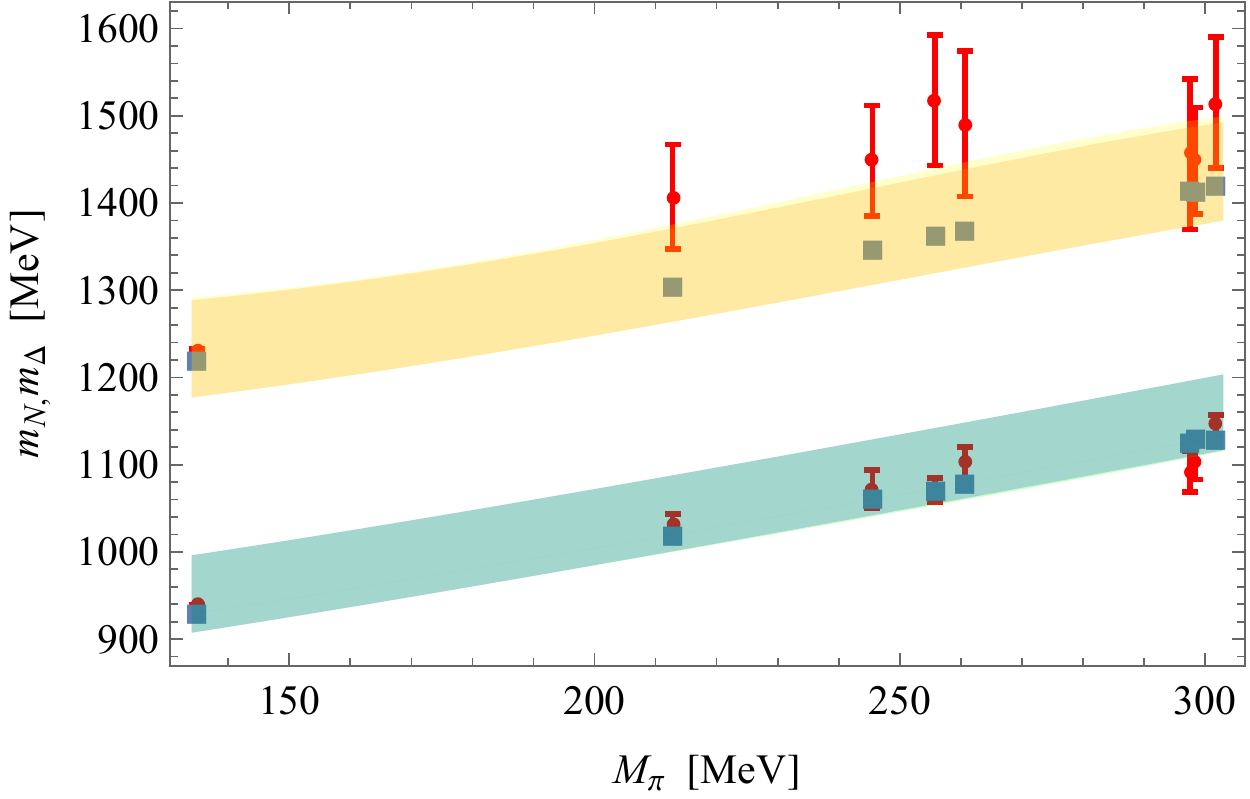,width=0.49\textwidth,angle=0} 
		\caption{From Ref. \cite{Fernando:2018jrz}; Left panel: summary of the   determinations of $\sigma_{\pi N}$  from $\pi N$ scattering (blue), from LQCD  (red), and from this work showing the combined  fit and   theoretical error. Right panel: $N$  and $\Delta$ masses from Fit 2 of Table \ref{Table:fits}: physical and LQCD masses from  \cite{Alexandrou:2016xok}. The squares are the results from the fit and the error bands correspond to 68\% confidence interval. Note: The references given in the left panel can be found in Ref. \cite{Fernando:2018jrz}}
		\label{Fig:Sigma_terms_figure}
	\end{center}
\end{figure}  
It is important to stress that  the resulting LECs  and the respective errors are  natural have natural size. More accurate LQCD results and, as emphasized later, with smaller $m_s$ would help determine how reliable is the effective theory is. Indeed, the behavior of $\sigma_{sN}$ as a function of $m_s$ shown in Figs. \ref{fig:1} and \ref{fig:sigmaSigmamq} indicates that the physical value of $m_s$ is too large for trusting the result obtained here. As discussed later, a qualitative picture in the limit of a heavy $m_s$ suggests a small value for $\sigma_s$ vanishing in the large quark mass limit. For the purpose of giving a constraint  of the contribution of  $\sigma_{sN}$ in Eqn. \ref{eq:sigma-relation-1} the analysis carried out here seems nonetheless adequate.
More details on extracting sigma terms for the Nucleon can be found in \cite{Fernando:2018jrz}. The fit gives an estimate  for $\sigma_{sN}$, which as discussed below is not credible, and should only be taken as an estimate of its magnitude for the purpose of determining $\sigma_{\pi N}$. As expected the results for the $\Delta$'s  $\sigma$ terms are very similar to those of the nucleon (they also have a small imaginary part due to the width of the $\Delta$).
A summary of the present status of $\sigma_{\pi N}$ determinations is  displayed in Fig. \ref{Fig:Sigma_terms_figure}.

\subsection{Dependencies on quark masses}

For $N$ and $\Delta$ the dependency of their masses on $\hat m$ is quite smooth up to $M_\pi\sim 300$ MeV (Fig.  \ref{Fig:Sigma_terms_figure}) .  In the case of the hyperons the dependency is less smooth the larger the magnitude of the strangeness (Fig. \ref{fig:LambdaSigmaMassEvolution}). The first indication of significant curvature appears in the $\sigma$ terms as the corresponding quark mass reaches a value of about 80 MeV, or about 300 MeV for the corresponding GB masses, as illustrated by Figs. \ref{fig:LambdaSigmaMassEvolution}.   This manifests itself in   curvature of the baryon masses with respect to quark masses but much less pronounced, consequence of Eq. \ref{eq:DeltamB}. 
\begin{figure}[h!!]
	\begin{center}
		\epsfig{file=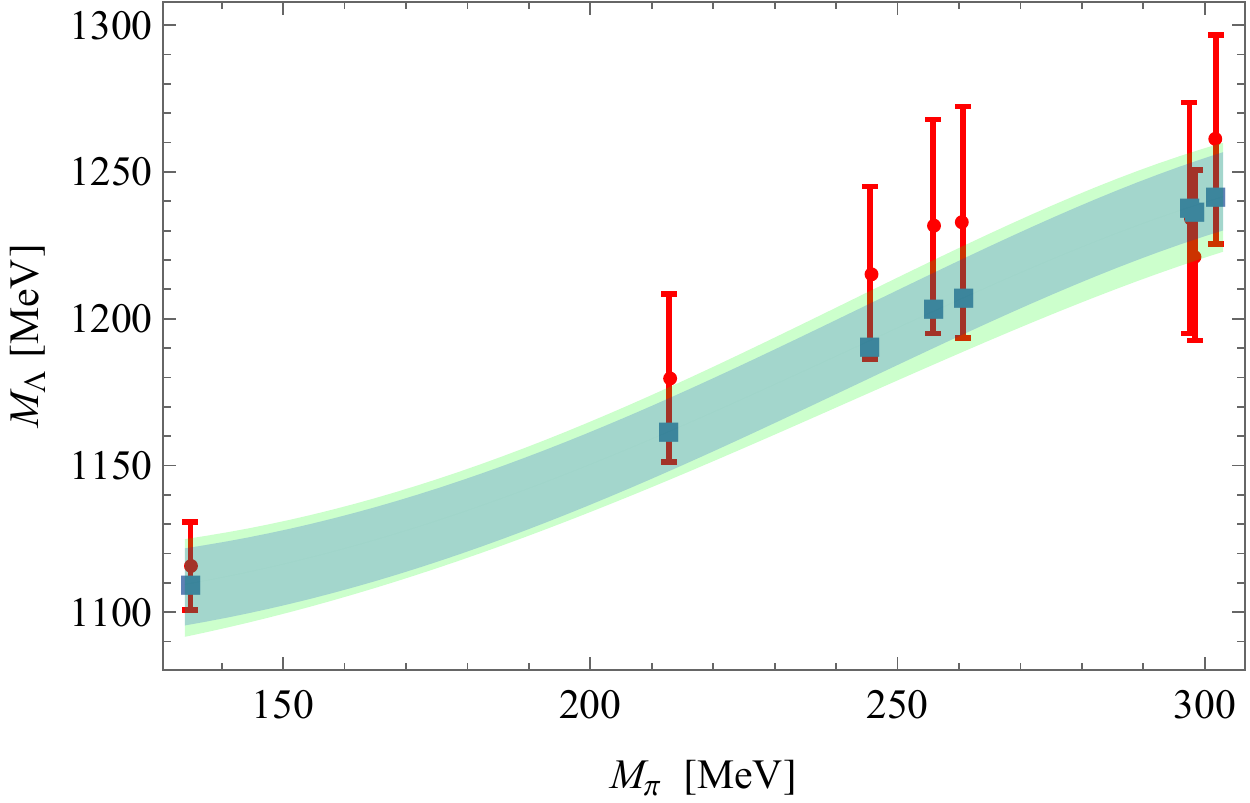,width=0.48\textwidth,angle=0} ~\epsfig{file=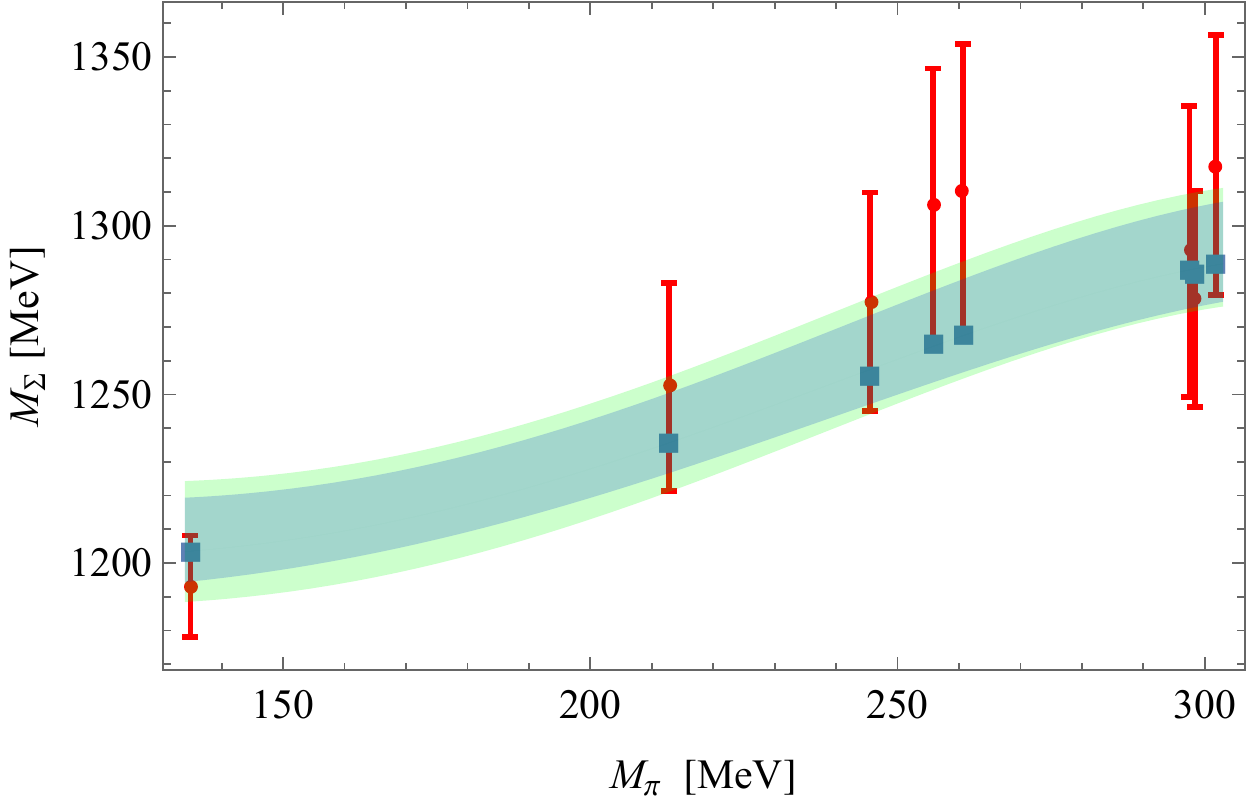,width=0.48\textwidth,angle=0} 
		\caption{ Evolution of $\Lambda$ (left panel) and $\Sigma$ baryon masses with $M_\pi$ at fixed $m_s$. The LQCD results in red are from Ref. \cite{Alexandrou:2016xok}}
		\label{fig:LambdaSigmaMassEvolution}
	\end{center}
\end{figure}
One can therefore estimate the range of quark masses for which the effective theory can describe baryon masses. 
For hadrons with a single heavy quark one can use results from HQET  to determine the hadron mass as a function of the heavy quark mass  \cite{Brambilla:2017hcq}, for which there would be a corresponding $\sigma$ term. Provided a definition of the heavy quark mass, the corresponding $\sigma$ term  will be, up to additive corrections determined by the scale of QCD, roughly proportional to the heavy quark mass with a slope close to unity.  In general, the slope is expected to scale roughly as proportional to the number of heavy quarks, and thus one can use this to give a rough estimate for the limit where the effective low energy theory ceases to describe a $\sigma$ term. For small quark masses the slope of the $\sigma$ term is much larger than it would be for the corresponding quark having a very large mass. The behavior of the $\sigma$ terms shown in Fig. \ref{fig:LambdaSigmaMassEvolution} illustrate  the natural tendency to a reduced slope as the quark mass increases.
One could therefore use the criterion that when the   slope calculated in the effective low energy theory reaches a value close to the one corresponding to the large quark mass limit, the   theory cannot  further be  trusted, representing this also the onset of  its  failure  for describing the hadron mass itself.  The analysis shown here indicates that this occurs  for  the relevant GB masses above 300 MeV or so. For this reason it would be very useful to have LQCD results where $m_s$ is taken to be smaller than  in present calculations, in order to assess more accurately the issue.  \begin{figure}[h!!]
	\begin{center}
		\epsfig{file=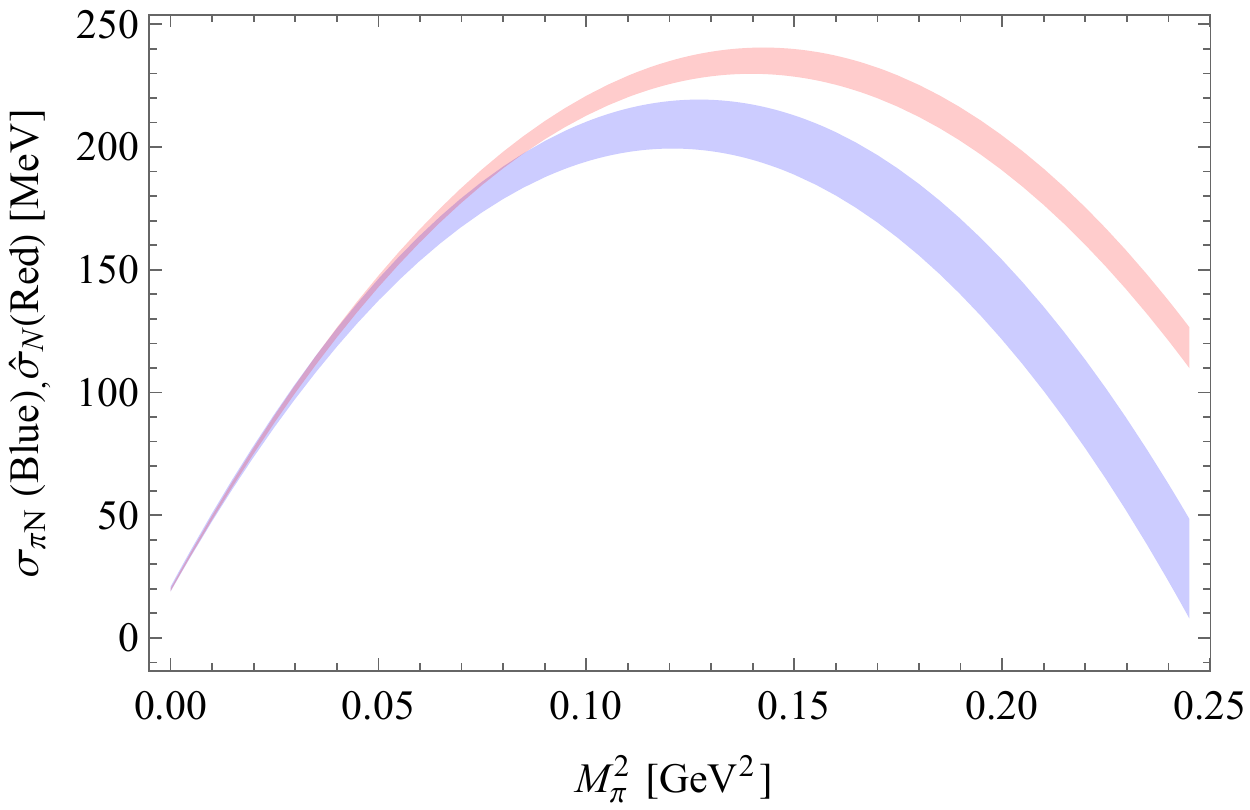,width= 0.360\textwidth,angle=0} \hspace*{1cm}~~~~~~\epsfig{file=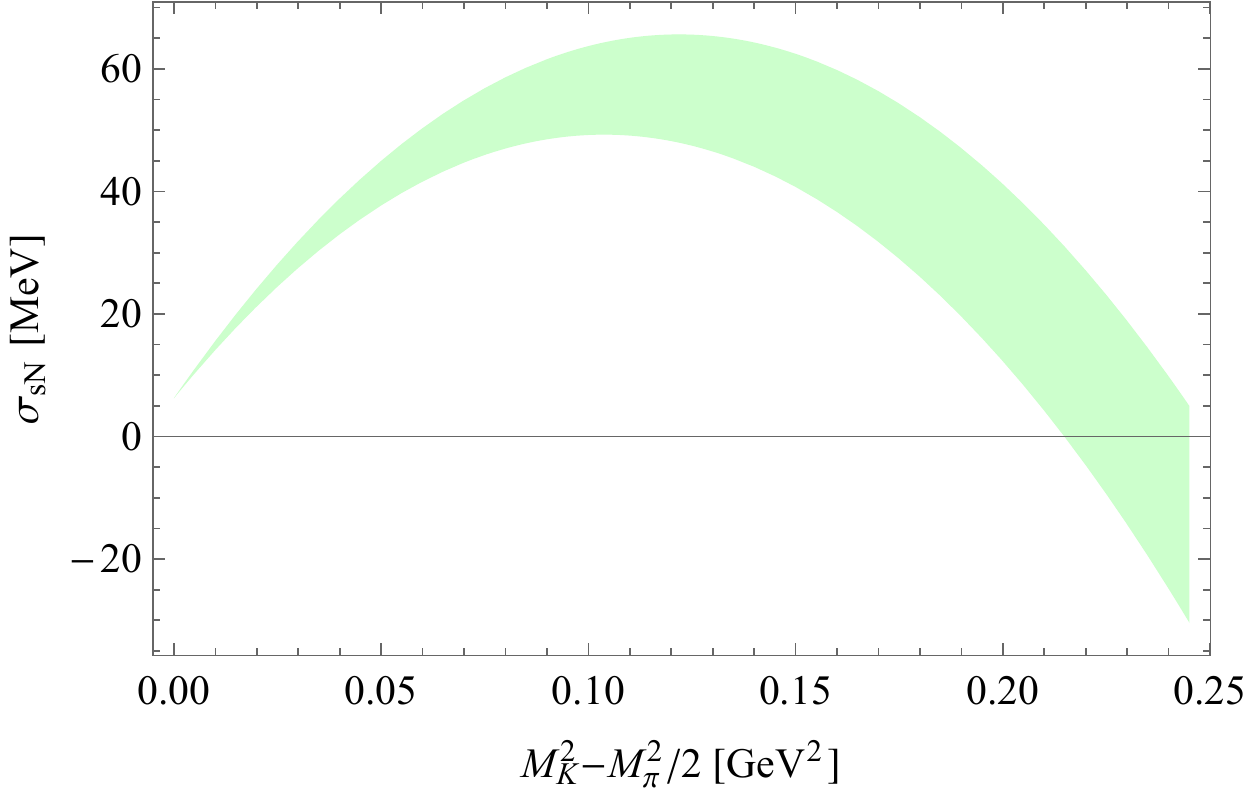,width= 0.36\textwidth,angle=0} 
		\epsfig{file=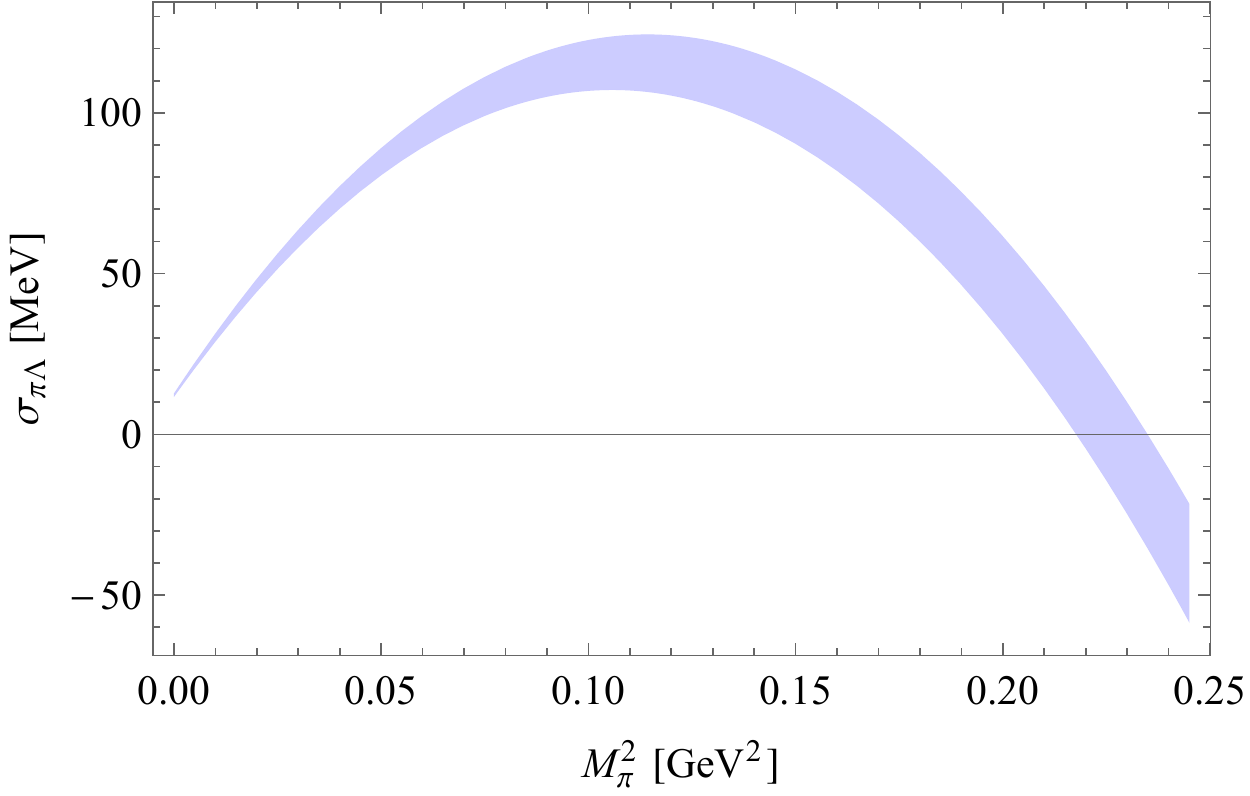,width= 0.36\textwidth,angle=0} \hspace*{1cm}~~~~~~\epsfig{file=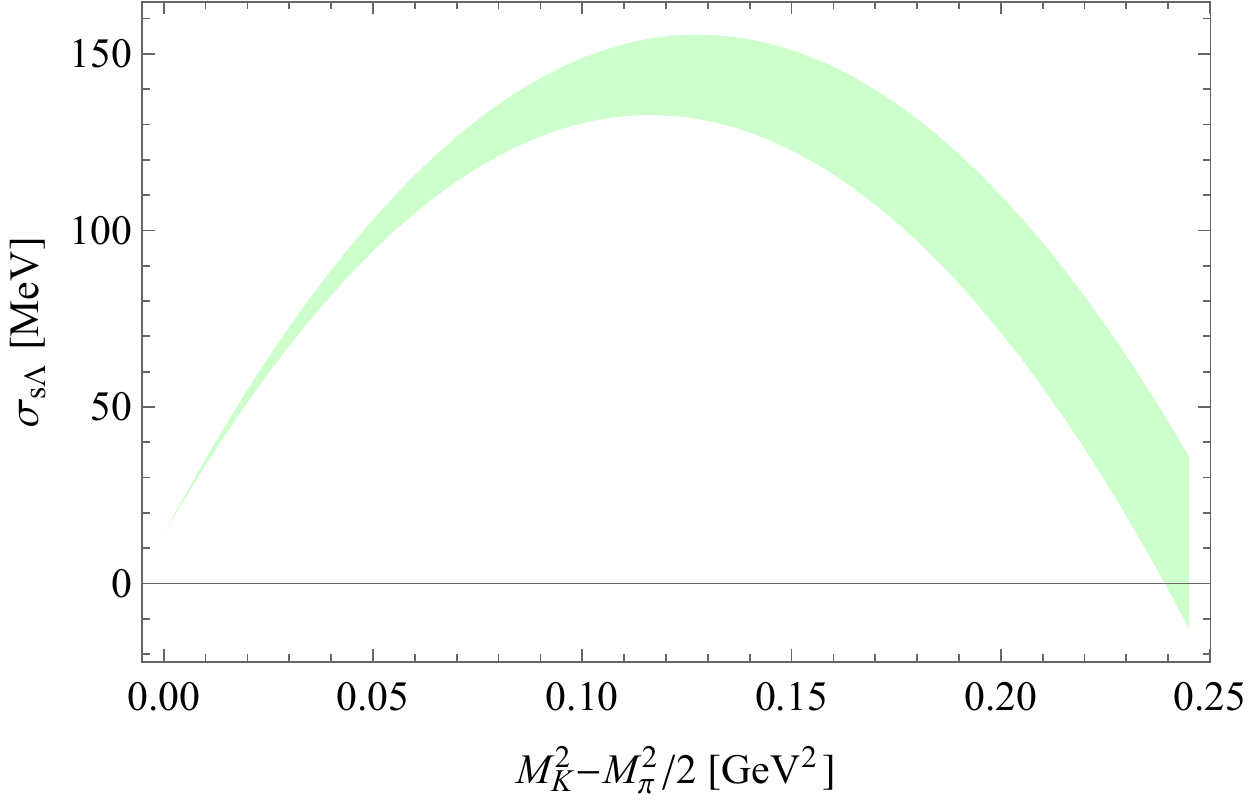,width= 0.36\textwidth,angle=0} 
		\epsfig{file=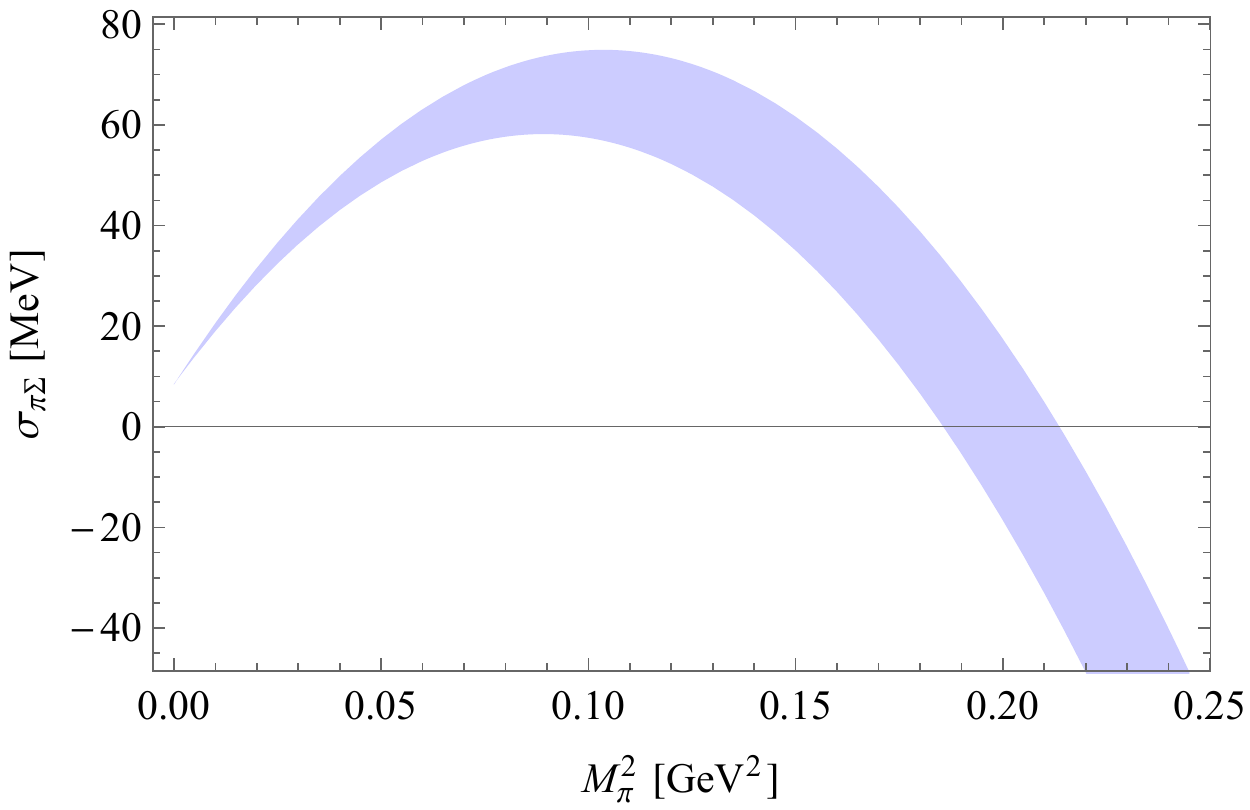,width= 0.36\textwidth,angle=0}\hspace*{1cm}~~~~~~\epsfig{file=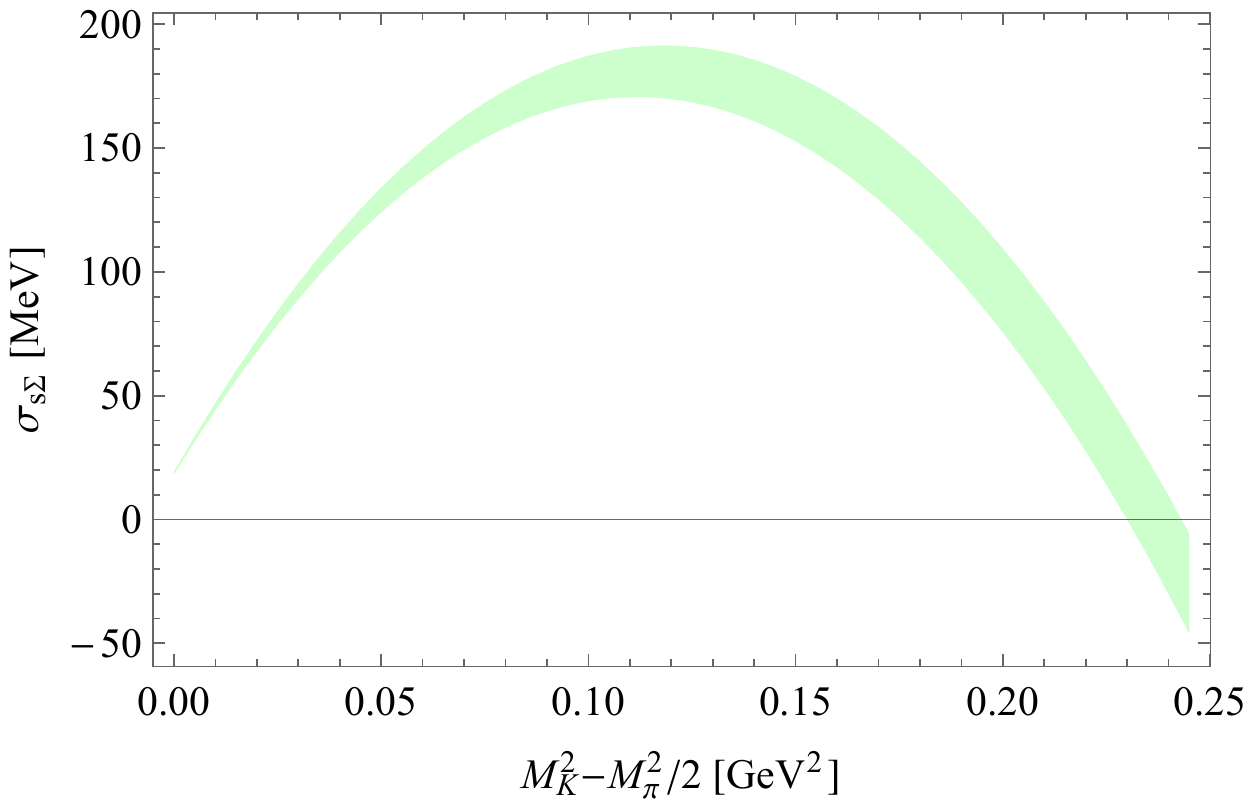,width= 0.36\textwidth,angle=0} 
 		\caption{$\sigma$ terms as function of quark masses.  In the left panels  $m_s$ is kept fixed, and in the right panels $\hat m$ is kept fixed. }
	\label{fig:sigmaSigmamq}
	\end{center}
\end{figure}

\section{Summary}
The determination of $\sigma$ terms through the Feynman-Hellmann theorem has its challenges. In principle a good knowledge of baryon masses for varying quark masses would be sufficient, but that knowledge as obtained from LQCD results is still not accurate enough to deliver values for $\sigma_{\pi N}$ with a precision near that  obtained from the analysis of $\pi N$ scattering. Another approach using   \BChPTNc in $SU(3)$  and its predictions for $\Delta_{GMO}$ and $\Delta \sigma_{8 N}$ as described in this note is potentially affected by the fact that $m_s$ is  too large for the result to be considered accurate. It is however interesting that an extraction of $\sigma_{\pi N}$ using that approach and the LQCD results agree very well. A result for $\sigma_{\pi N}=69(10)$ MeV results from those analyses, consistent with the larger values obtained from $\pi N$ scattering.
It should be emphasized that a similar analysis using ordinary $BChPT$ with only the octet baryons completely fails in that respect.  We also learn that the description of strangeness $\sigma$ terms fails for the physical value of $m_s$, and thus, one would need   LQCD  results with reduced values of $m_s$ to understand more precisely the range where effective theories can describe them: it looks like the for the effective theory to be able to reliably describe $\sigma$ terms in $SU(3)$ would require $M_K\leq 350$ MeV.

\section{Acknowledgements}
The authors would like to thank Jos\'{e} Manuel Alarc\'{o}n for earlier collaboration and  useful discussions.  This work was supported by DOE Contract No. DE-AC05-06OR23177 under which JSA operates the Thomas Jefferson National Accelerator Facility, 
and by the  National Science Foundation through grants  PHY-1307413 and PHY-1613951.

\bibliography{Refs}

\end{document}